\documentclass[twocolumn,
aps,prd,nofootinbib,
superscriptaddress,
showkeys,
tightenlines,
amsmath,
amssymb
]{revtex4}

\usepackage[english]{babel}
\usepackage{graphicx}
\usepackage{epsf}
\usepackage{pstricks}

\makeindex
\begin{document}

\title{Recent and future experimental evidences for exotic mesons in hard reactions}

\author{\underline{I.V.~Anikin}}

\affiliation{Bogoliubov Laboratory of Theoretical Physics, JINR, 141980 Dubna, Russia}

\author{O.V.~Teryaev }

\affiliation{Bogoliubov Laboratory of Theoretical Physics, JINR, 141980 Dubna, Russia}

\author{B.~Pire}

\affiliation{CPHT, {\'E}cole Polytechnique, CNRS, 91128 Palaiseau Cedex, France}

\author{L.~Szymanowski}

\affiliation{CPHT, {\'E}cole Polytechnique, CNRS, 91128 Palaiseau Cedex, France\\
University of Li{\`e}ge, B-4000 Li{\`e}ge, Belgium\\
Soltan Institute for Nuclear Studies, Warsaw, Poland}

\author{S.~Wallon}

\affiliation{LPT, Universit{\'e} Paris-Sud, CNRS, 91405 Orsay, France }

\vspace{0.2cm}
\begin{abstract}

\noindent
The QCD analysis of the recent experimental data (L3@LEP) 
of
the hard exclusive $\rho\rho$ production in two photon collisions shows that these data 
can be understood as a signal for the existence of an exotic isotensor resonance with 
a mass around $1.5\, {\rm GeV}$. 
We also argue that hard exclusive reactions are a powerful tool for 
an experimental 
study of exotic hybrid mesons with $J^{PC}=1^{-+}$.

\end{abstract}
\keywords{Hard reaction, exotic mesons, hybrids, GPDs, GDAs.}
\maketitle

\section{Introduction}

\noindent
Exclusive reactions $\gamma^*\gamma\to A + B $ which may be accessed in
$e^+ e^-$ collisions have been shown \cite{DGPT} to have a partonic
interpretation in the kinematical region of large virtuality of one photon and
of small center of mass energy. First data on the
$\rho^0 \rho^0$ channel at
LEP have been published \cite{L3Coll1} and analyzed \cite{APT}, showing
the compatibility of the QCD leading order analysis with experiment at quite modest
values of $Q^2$.

We first focus on the comparison of processes $\gamma^* \gamma
\to \rho^0\rho^0$
and $\gamma^* \gamma \to \rho^+\rho^-$ in the context of searching
an exotic isospin $2$ resonance decaying into two $\rho$ mesons; such
channels have recently been studied at LEP by the L3 collaboration 
\cite{L3Coll1,L3Coll2}. A
related study for photoproduction \cite{Rosner}
raised the problem of $\rho^0\rho^0$ enhancement with respect to
$\rho^+\rho^-$ at low energies.
One of the solutions of this problem was based on the prediction \cite{Achasov0} and
further exploration \cite{Achasov} of
the possible existence
of isotensor state, whose interference with the isoscalar state is constructive for neutral mesons
and destructive for charged ones.
The crucial property of such an exotic state is the absence of $\bar q q$ wave function at any
momentum resolution. In other words, quark-antiquark component is absent both in its non-relativistic
description and
at the level of the light-cone distribution amplitude. 
An isotensor state on the light cone corresponds to the twist $4$ or higher and
its  contribution is thus power suppressed at large $Q^2$.
This is supported by the mentioned L3 data, where
the high $Q^2$ ratio of the cross sections of charged and neutral mesons production
points out an isoscalar state.

In \cite{isotenAPT}, both perturbative and non-perturbative ingredients of QCD factorization for the
description of an isotensor state have been studied.
Namely, the twist $4$ coefficient function has been calculated and the 
non-perturbative matrix elements has been extracted from L3 data.
The analysis of \cite{isotenAPT} is compatible with the existence of an 
isotensor exotic meson with a mass around $1.5$  GeV.

Other exotic objects which can be investigated in hard reactions are 
the mesons with exotic quantum numbers, for example, with $J^{PC}=1^{-+}$. These mesons,
which have been called hybrids, cannot be described within the usual quark model. 
We have studied the hybrid mesons in the deep exclusive electroproduction processes
which is well described in the framework of the collinear
approximation where generalized parton distributions (GPDs) and distribution
amplitudes describe the nonperturbative parts of a factorized amplitude.
In \cite{APSTW1}, we showed that contrarily to naive expectations, the
amplitude for the electroproduction of an exotic meson with $J^{PC} = 1^{-+}$ may be
written in a very similar way as the amplitude for non-exotic vector meson electroproduction.
The main observation of our work was that the
quark-antiquark correlator on the light cone includes a gluonic component due to gauge
invariance and leads to a leading twist hybrid light-cone distribution amplitude.
We also studied the hybrid meson  as a resonance in the reaction $e\,p\to e\,p\,(\pi^0\eta)$.
The first experimental investigation of the hybrid with $J^{PC}=1^{-+}$ as the resonance in
$\pi^-\eta$ mode was implemented by the Brookhaven collaboration E852 \cite{E852}.
Present candidates for the hybrid states with $J^{PC}=1^{-+}$ include $\pi_{1} (1400)$ which is
mostly seen through its $\pi\eta$ decay and $\pi_{1} (1600)$ which is seen through its
$\pi \eta' $ and $\pi \rho $ decays \cite{RPP}.

\section{Searching isotensor exotic mesons in $\gamma^* \gamma \to \rho\rho$ process}

\noindent
The reaction which we study here is \\
$e(k)+e(l)\to e(k^{\prime})+e(l^{\prime})+\rho(p_1)+\rho(p_2)$,
where $\rho$ stands for the triplet $\rho$ mesons;
the initial  electron $e(k)$ radiates
a hard virtual photon with momentum $q=k-k^{\prime}$,
with $q^2=-Q^2$ quite large. This means that the scattered electron  $e(k^{\prime})$
is tagged.
To describe the given reaction, it is useful to consider the subprocess 
$e(k)+\gamma(q^{\prime})\to e(k^{\prime})+\rho(p_1)+\rho(p_2)$.
Regarding the other photon momentum $q^{\prime}=l-l^{\prime}$,
we assume that, firstly, its momentum is almost collinear to the electron
momentum $l$ and, secondly, that $q^{\prime \, 2}$ is approximately
equal to zero, which is a usual approximation when the second lepton
is untagged.
In two $\rho$ meson production, we are interested in the channel where
the resonance corresponds to the exotic isospin, {\it i.e} $I=2$, and usual
$J^{PC}$ quantum numbers. The $J^{PC}$ quantum numbers are not essential for our study.
Because the isospin $2$ state has only a projection on the four quark correlators,
the study of mesons with the isospin $2$ can help to throw light upon the
four quark states. We thus, together with the mentioned reactions,
study the following processes:
$e(k)+e(l)\to e(k^{\prime})+e(l^{\prime})+ R(p)$
and $e(k)+\gamma(q^{\prime})\to e(k^{\prime})+R(p)$,
where meson $R(p)$ possesses isospin $I=2$.
The amplitudes
for $\rho^0\rho^0$ and $\rho^+\rho^-$ productions can be written in the form of
the decomposition over the
amplitudes associated with the two and four quark correlators.
The amplitudes corresponding to $\rho^+\rho^-$
production are not independent and can be expressed through the corresponding amplitudes
of $\rho^0\rho^0$ production (see, \cite{isotenAPT}). 
Note that, in our case, only isospin $0$ and $2$ cases are relevant due to
the positive $C$-parity of the initial and final states. 

The differential cross sections
$d\sigma_{ee\to ee\rho\rho}/dQ^2\, dW^2$ for both the $\rho^0\rho^0$ and
$\rho^+\rho^-$  channels contain a
number of unknown phenomenological parameters, which are intrinsically 
related to non perturbative
quantities encoded in the generalized distribution amplitudes.
We implemented a fit of these phenomenological parameters
in order to get a good description of experimental data.
We have obtained the following set of parameters for fitting:
the masses and widths of isoscalar and isotensor resonances 
$M_{R^0}$, $\Gamma_{R^0}$, $M_{R^2}$, $\Gamma_{R^2}$, and the phenomenological parameters 
${\bf S}^{I=0,I_3=0}_2$, ${\bf S}^{I=0,I_3=0}_4$, ${\bf S}^{I=2,I_3=0}_4$ 
which are related with the corresponding 
matrix elements of twist $2$ and twist $4$ operators (for more details, see
\cite{isotenAPT}).
Studying both the $W$ dependence and the $Q^2$ dependence of 
$\rho^0\rho^0$ and $\rho^+\rho^-$ production cross sections, it turned out
that the best description is reached at
$M_{R^2}=1.5\, {\rm GeV}$, $\Gamma_{R^2}=0.4 \,{\rm GeV}$, ${\bf S}^{I=0,I_3=0}_2=0.12 \,{\rm GeV}$,
${\bf S}^{I=0,I_3=0}_4=0.006 \,{\rm GeV}$, ${\bf S}^{I=2,I_3=0}_4=0.018 \,{\rm GeV}.$

\noindent
Thus, the fitting of LEP data based on the QCD factorization of the amplitude into a hard subprocess and
a generalized distribution amplitude is consistent with the existence of an isospin $I=2$
exotic meson \cite{Achasov0,Achasov} with a mass in the vicinity of $1.5\, {\rm GeV}$
and a width around $0.4\, {\rm GeV}$. The contributions of such an exotic meson in the two
$\rho$ meson production cross sections are directly associated with
some twist $4$ terms that we have identified. 

\noindent
Analysing the $Q^2$ dependence, we can see that due to the presence of a twist $4$ amplitude
and its interference with the leading twist $2$ component, the $\rho^0\rho^0$ cross section at
 small $Q^2$ is a few times higher than the
$\rho^+\rho^-$ cross section.
While for the region of large $Q^2$ where any higher twist effects are
negligible the $\rho^0\rho^0$ cross section is less than the $\rho^+\rho^-$ cross section by
the  factor $2$.

\section{Exotic hybrid meson search in hard electroproduction}

\noindent
Within quantum chromodynamics, hadrons are described in terms of quarks, anti-quarks and
gluons. The usual, well-known, mesons are supposed to contain quarks and
anti-quarks as valence
degrees of freedom while gluons play the role of carrier of interaction,
{\it i.e.} they remain hidden in a background.
On the other hand, QCD does not prohibit the existence of the explicit gluonic degree of freedom
in the form of a vibrating flux tube, for instance. The states where the $q\bar q g$ and
$gg$ configurations are dominating, hybrids and glueballs,
are of fundamental importance to understand the dynamics of quark confinement and the
nonperturbative sector of quantum chromodynamics \cite{Close}.

In \cite{hybrid}, we proposed to study the exotic hybrid meson by means of its deep exclusive
electroproduction, {\it i.e.}
$e(k_1)\, + \, N(p_1)\,\to\,e(k_2)\,+ H(p)\,+\,N(p_2)$,
where we concentrate on the subprocess:
$\gamma^*_L(q)\, + \, N(p_1)\,\to\, H_L(p)\,+\,N(p_2)$
when the baryon is scattered at small angle.
This process is a hard exclusive reaction due to the transferred momentum $Q^2$
is  large ( Bjorken regime).
Within this regime where a factorization theorem is valid,  at
the leading twist level, the crucial point is to properly define the hybrid meson distribution
amplitude. 
The Fourier transform of the hybrid meson --to--vacuum matrix element of
the bilocal vector quark operator may be written as
\begin{eqnarray}
&&\langle H_L(p,0)| \bar \psi(-z/2)[-z/2;z/2]\gamma_\mu
\psi(z/2) | 0 \rangle 
\nonumber\\
&&=i f_H M_H e^{(0)}_{L\,\mu}
\int\limits_0^1 dy e^{i(\bar y - y)p\cdot z/2} \phi^{H}_L(y)
\label{hmeW},
\end{eqnarray}
where $e^{(0)}_{L\,\mu}=(e^{(0)}\cdot z)/(p\cdot z) p_\mu$
and $\bar y=1-y$ and $H$ denotes the isovector triplet of hybrid mesons;
$f_H$ denotes a dimensionful coupling constant of the
hybrid meson, of the order of $50 $ MeV, so that $\phi^H_L$ is 
dimensionless.
$\phi^H_L$ is asymptotically equal to
\begin{eqnarray}
\label{approxH}
\phi_{L}^{H}(u)=30 u (1-u)(1-2u).
\end{eqnarray}

We have calculated 
the leading twist contribution to exotic hybrid meson with $J^{PC}=1^{-+}$ electroproduction
amplitude in the deep exclusive region.  The resulting order of magnitude
is somewhat smaller than the $\rho$ electroproduction but
similar to the $\pi$ electroproduction. The obtained cross section is
sizeable and should be measurable at dedicated experiments at JLab, Hermes or Compass.

\noindent
We made a systematic comparison with the non-exotic vector meson production.
To take into account NLO corrections, the differential cross-sections for these processes have
been computed using the BLM prescription for the renormalization scale.
In the case of $\rho$ production, our estimate is not far from a previous one  which
took into account kinematical higher twist corrections.

\noindent
We have also discussed in detail the $\pi\eta$ mode corresponding to the $\pi _{1}(1400)$
candidate in the reaction $e\, p\to e\,p\,\pi^0\eta$.
We have calculated an angular asymmetry  implied by charge conjugation properties
and got a sizeable hybrid effect which may be experimentally checked.
We also considered the process $\gamma^*\gamma\to H\to\pi\eta$ which may be accessible in 
$e^+ e^-$ collision.

\noindent
In the region of  small $Q^2$  higher twist contributions should be carefully studied and
included. Note that they have already been considered in the case of deeply
virtual Compton scattering \cite{APT} where their presence was dictated by gauge invariance, and
for transversely polarized vector mesons \cite{AT} where the leading twist component vanishes.

\section*{Acknowledgments}
This work is partly supported by the French-Polish 
scientific agreement Polonium, the Polish Grant 1 P03B 028 28; the Fonds 
National de la Recherche Scientifique (FNRS, Belgium); the ECO-NET 
program, contract 12584QK;
the Joint Research Activity "Generalised Parton Distributions" of the 
european I3 program Hadronic Physics, contract RII3-CT-2004-506078;
RFFI Grant 06-02-16215.

\end{document}